# Effect of the Earth's time-retarded transverse Gravitational field on the motion of the Moon

## (version 10)

## J. C. Hafele[1]

Classical Newtonian gravitational theory does not satisfy the causality principle because it is based on instantaneous action-at-a-distance. A causal version of Newtonian theory for a large rotating sphere is derived herein by time-retarding the distance between interior circulating point-mass sources and an exterior field-point. The resulting causal theory explains exactly the flyby anomaly reported by NASA scientists in 2008. It also explains exactly an anomalous decrease in the Moon's orbital speed. No other known theory can make both of these claims.

## 1. Introduction

As reported in 2008 by Anderson *et al.*,[1] NASA has found anomalous orbital-energy changes during six spacecraft flybys of the Earth. The reported speed-changes range from +13.28 mm/s for the NEAR flyby to -4.67 mm/s for the Galileo-II flyby. Anderson *et al.* state in their abstract; "These anomalous energy changes are consistent with an empirical prediction formula which is proportional to the total orbital energy per unit mass and which involves the incoming and outgoing geocentric latitudes of the asymptotic spacecraft velocity vectors." If the calculated speed-change is designated by $\delta v_{emp}$, the empirical prediction formula can be expressed as follows.

$$\delta v_{emp} = \frac{2v_{Eq}}{c} v_{in} \left(\cos(\lambda_{in}) - \cos(\lambda_{out})\right) = -\frac{2v_{Eq}}{c} v_{in} \int_{t_{in}}^{t_{out}} \sin(\lambda(t)) \frac{d\lambda}{dt} dt \quad , \quad (1.1)$$

where $v_{Eq}$ is the Earth's equatorial rotational surface speed, c is the vacuum speed of light, $v_{in}$ is the initial asymptotic inbound speed, $\lambda_{in}$ is the asymptotic inbound geocentric latitude, and $\lambda_{out}$ is the asymptotic outbound geocentric latitude. The coordinate time for the spacecraft in its trajectory is t, so that $\lambda_{in}=\lambda(t_{in})$ and $\lambda_{out}=\lambda(t_{out})$. If $d\lambda/dt=0$, then $\delta v_{emp}=0$.

The following is a direct quote from the report by Anderson *et al.* (The ODP means the Orbit Determination Program.)

> Lämmerzahl *et al.* [11] studied and dismissed a number of possible explanations for the Earth flyby anomalies, including Earth atmosphere, ocean tides, solid Earth tides, spacecraft charging, magnetic moments, Earth albedo, solar wind, coupling of Earth's spin with rotation of the radio wave, Earth gravity, and relativistic effects predicted by Einstein's theory. All of these potential sources of systematic error, and more, are modeled in the ODP. None can account for the observed anomalies.

The objective of this article is to derive a new causal version of classical acausal Newtonian theory, and show that this new version is able to produce exact agreement with the speed-changes reported by Anderson *et al.* This new version proposes a simple correction that converts Newton's acausal theory into a causal theory. The resulting causal theory contains a new, previously overlooked, time-retarded transverse component, designated $\mathbf{g}_{trt}$. The resulting total gravitational field consists of two components, the standard well-known classical radial component, $\mathbf{g}_r$, and the new time-retarded transverse vortex component, $\mathbf{g}_{trt}$.

The formula for the signed magnitude of $\mathbf{g}_{trt}$ will be shown in Section 2 to be

$$g_{trt}(\theta) = -G \frac{I_E}{r_E{}^4} \frac{v_{Eq}}{c_g} \frac{\Omega_\phi(\theta) - \Omega_E}{\Omega_E} \cos^2(\lambda(\theta)) PS(r(\theta)) \;, \tag{1.2}$$

where G is the gravity constant, $I_E$ is the Earth's spherical moment of inertia, $r_E$ is the Earth's spherical radius, $\Omega_E$ is the Earth's spin angular speed, $c_g$ is the speed of the Earth's gravitational field, $\theta$ is the spacecraft's parametric polar coordinate angle in the plane of the trajectory, $\Omega_\theta \equiv d\theta/dt$ is the spacecraft's angular speed, $\Omega_\phi$ is the azimuthal $\phi$-component of $\Omega_\theta$, $\lambda$ is the spacecraft's geocentric latitude, $r$ is the spacecraft's geocentric radial distance, and PS(r) is an inverse-cube power series representation for a triple integral over the Earth's volume. If the magnitude is negative ($\Omega_\phi > \Omega_E$), the vector field component $\mathbf{g}_{trt}$ is directed towards the east.

The empirical prediction formula (*cf.* Eq. (1.1)) shows that the speed-change must be in the $\lambda$-component of the spacecraft's velocity, $\mathbf{v}_\lambda$. The magnitude for the $\lambda$-component is defined by

$$v_\lambda = r_\lambda \frac{d\lambda}{dt} = r_\lambda \frac{d\lambda}{d\theta} \frac{d\theta}{dt} = r_\lambda \Omega_\theta \frac{d\lambda}{d\theta} \;, \tag{1.3}$$

where $r_\lambda$ is the $\lambda$-component of r. The velocity component, $\mathbf{v}_\lambda$, is orthogonal to $\mathbf{g}_{trt}$. Consequently, $\mathbf{g}_{trt}$ cannot directly change the <u>magnitude</u> of the spacecraft's velocity.

However, a hypothesized "induction-like" field, designated $\mathbf{F}_\lambda$, can be directed perpendicularly to $\mathbf{g}_{trt}$ in the $\mathbf{v}_\lambda$-direction. Assume that $\mathbf{F}_\lambda$ is proportional to the time derivative of $\mathbf{g}_{trt}$, $kd\mathbf{g}_{trt}/dt$. This induction-like field causes a small change in the spacecraft's speed. The reciprocal of k, $v_k \equiv 1/k$, becomes an adjustable parameter, herein called the "induction speed". The formula for the magnitude of $\mathbf{F}_\lambda$ will be shown in Section 3 to be

$$F_\lambda(\theta) = \frac{v_{Eq}}{v_k} \frac{r_E}{r(\theta)} \int_0^{\theta} \frac{r(\theta)}{r_E} \frac{\Omega_\theta(\theta)}{\Omega_E} \frac{1}{r_E} \frac{dr}{d\theta} \frac{dg_{trt}}{d\theta} d\theta \;. \tag{1.4}$$

The "anomalous" time rate of change in the spacecraft's kinetic energy is given by the dot product, $\mathbf{v} \cdot \mathbf{F}_\lambda$. It will be shown in Section 4 that the calculated asymptotic speed-change, $\delta v_{trt}$, is given by

$$\delta v_{trt} = \delta v_{in} + \delta v_{out} \;, \tag{1.5}$$

Table I. Listing of the observed speed change, $\delta v_{obs}$, the calculated speed change, $\delta v_{trt}$, the ratio used for the speed of gravity, $c_g/c$, the required value for the induction speed, $v_k/v_{Eq}$, and the eccentricity, $\varepsilon$, for each of the six Earth flybys reported by Anderson et al.[1] The two high-precision flybys are marked by *.

| flyby | NEAR* | GLL-I | Rosetta* | M'GER | Cassini | GLL-II |
|-------|-------|-------|----------|-------|---------|--------|
| $\delta v_{obs}$(mm/s) | +13.46 ±0.01 | +3.92 ±0.30 | +1.80 ±0.03 | +0.02 ±0.01 | -2 ±1 | -4.6 ±1 |
| $\delta v_{trt}$(mm/s) | +13.46 ±0.01 | +3.92 ±0.30 | +1.80 ±0.03 | +0.02 ±0.01 | -2 ±1 | -4.6 ±1 |
| $c_g/c$ | 1.060 ±0.001 | 1.0 ±0.1 | 1.06 ±0.02 | 1.0 ±0.5 | 1.0 ±0.5 | 1.0 ±0.2 |
| $v_k/v_{Eq}$ | 4.130 ±0.003 | 16 ±2 | 7.74 ±0.13 | 10 ±5 | 23 ±12 | 16 ±4 |
| $\varepsilon$ | 1.8142 | 2.4731 | 1.3122 | 1.3596 | 5.8456 | 2.3186 |

where

$$\delta v_{in} = \delta v(\theta_{min}) \text{ and } \delta v_{out} = \delta v(\theta_{max}) \text{ ,} \qquad (1.6)$$

and

$$\delta v(\theta) = \frac{v_{in}}{2} \int_0^\theta \frac{r_\lambda F_\lambda}{v_{in}^2} \frac{d\lambda}{d\theta} d\theta \text{ .} \qquad (1.7)$$

The angles $\theta_{min}$ and $\theta_{max}$ are the minimum and maximum values for $\theta$. The initial speed $v_{in}=v(\theta_{min})$.

This "neoclassical" causal version for acausal Newtonian theory explains exactly the flyby anomaly.[2] Table I lists, for each of the six Earth flybys reported by Anderson et al., the observed speed change, $\delta v_{obs}$, the calculated speed change, $\delta v_{trt}$, the ratio that was used for the speed of gravity, $c_g/c$, the value for the induction speed that gives exact agreement with the observed speed-change, $v_k/v_{Eq}$, and the value for the eccentricity of the trajectory, $\varepsilon$. Notice in Table I that the required values for $v_k$ cluster between $4v_{Eq}$ and $23v_{Eq}$. Also notice that the two high-precision flybys, NEAR and Rosetta, put very stringent limits on the speed of gravity, $c_g$. If the "true" value for $v_k$ had been known with high precision, the two high-precision flybys would have provided first-ever measured values for the speed of the Earth's gravitational field.

In 1995, F. R. Stephenson and L. V. Morrison published a remarkable study of records of eclipses from 700 BC to 1990 AD.[3] They conclude (LOD means length-of-solar-day and ms cy$^{-1}$ means milliseconds per century):

(1) the LOD has been increasing on average during the past 2700 years at the rate of +1.70±0.05 ms cy$^{-1}$ ((+17.0±0.5)×10$^{-6}$ s per year),
(2) tidal braking causes an increase in the LOD of +2.3±0.1 ms cy$^{-1}$ ((+23±1)×10$^{-6}$ s per year),
(3) so that there is a non-tidal decrease in the LOD, numerically -0.6±0.1 ms cy$^{-1}$ ((-6±1)×10$^{-6}$ s per year).

Stephenson and Morrison state that the non-tidal decrease in the LOD probably is caused by post-glacial rebound. Post-glacial rebound decreases the Earth's moment of inertia, which increases the Earth's spin angular momentum, and thereby decreases the LOD. But post-glacial rebound cannot change the Moon's orbital angular momentum.

According to Stephenson and Morrison, tidal braking causes an increase in the LOD of $(23\pm1)\times10^{-6}$ s per year, which causes a decrease in the Earth's spin angular momentum, and by conservation of angular momentum causes an increase in the Moon's orbital angular momentum. It will be shown in Section 5 that tidal braking alone would cause an increase in the Moon's orbital speed of $(19\pm1)\times10^{-9}$ m/s per year, which corresponds to an increase in the radius of the Moon's orbit of $14\pm1$ mm per year.

Lunar laser ranging experiments have shown that the radius of the Moon's orbit is actually increasing at the rate of $38\pm1$ mm per year.[4] This rate for increase in the radius corresponds to an increase in the orbital speed of $(52\pm2)\times10^{-9}$ m/s per year. Clearly there is an unexplained or anomalous difference in the change in the radius of the orbit of $(-24\pm2)$ mm per year (38-14=24), and a corresponding anomalous difference in the change in the orbital speed of $(-33\pm3)\times10^{-9}$ m/s per year (52-19=33). This "lunar orbit anomaly" cannot be caused by post-glacial rebound, but it can be caused by the proposed neoclassical causal version of Newton's theory.

It will be shown in Section 6 that the neoclassical causal version produces a change in the Moon's orbital speed of $(-33\pm3)\times10^{-9}$ m/s per year if the value for the induction speed $v_k=(7\pm1)v_{Eq}$. The eccentricity for the Moon's orbit, ε=0.0549, indicates that it is a nearly circular closed orbit. Consequently, a new closed orbit case can be added to the open orbit flybys listed in Table I. A graph of the induction speed ratio, $v_k/v_{Eq}$, versus eccentricity, ε, Fig. 1, shows that the required value for the induction speed for the Moon is consistent with the required values for the induction speed for the six flyby anomalies. The least-squares linear regression line suggests that the induction speed may increase with increasing ε.

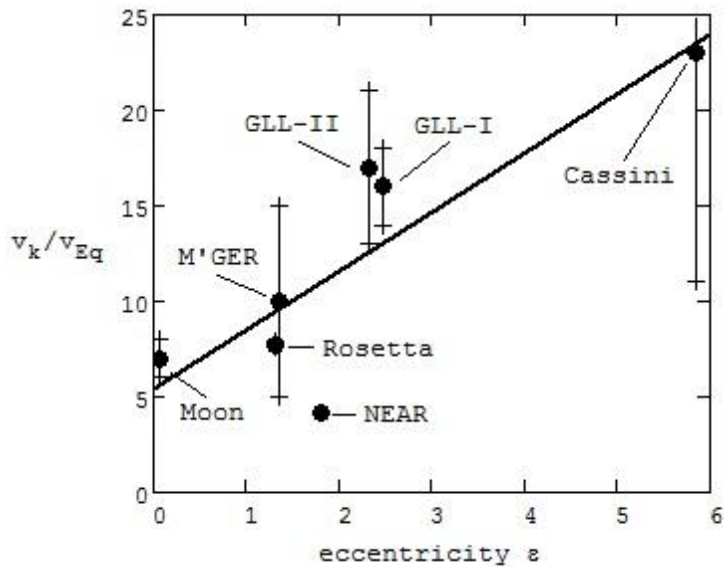

Figure 1. Required induction speed ratio (designated by •), $v_k/v_{Eq}$ ± the assigned uncertainty, versus eccentricity, ε. The upward increasing solid line is a least-squares linear-regression line with a slope of +3.1 and a y-axis intercept of +5.4.

In conclusion, the proposed neoclassical causal version for acausal Newtonian theory has passed seven tests: (1) explanation of the six flyby anomalies, and (2) explanation of the lunar orbit anomaly. It will be very difficult if not impossible for any other rational theory to pass all of these tests.

## 2. Time-retarded transverse gravitational field for a large rotating sphere

Consider a spacecraft that is in either a closed Keplerian orbit or an open orbit around a large spinning isotropic rigid sphere (a spherical approximation for the Earth). Numerical values for $M_E$, $r_E$, $\Omega_E$, $v_{Eq}$, and $I_E$ (cf. Eq. (1.2)) are listed in Appendix A. Let $\rho_E(r)$ be the radial mass-density distribution, and let $dm(r)$ be an infinitesimal interior circulating point-mass source. Let the spherical coordinates for dm be $(r,\lambda',\phi')$, where $r$ is the interior radial distance, $\lambda'$ is the interior geocentric latitude, and $\phi'$ is the interior azimuthal angle. Let $t'$ be the coordinate time for dm, where $t'$ is the retarded time.

Let $(X,Y,Z)$ be a nonrotating geocentric frame of reference with the $(X,Y)$ plane being the equatorial plane and the Z-axis being the axis of rotation for the sphere. Let $\lambda_p$ be the spacecraft's geocentric latitude at perigee. Let $(x,y,z)$ be the frame in which the $(x,y)$ plane is the plane of the orbit or trajectory. Let the x-axis coincide with the X-axis if $\lambda_p$=0. Let $r(\theta)$ be the exterior geocentric radial distance to the spacecraft, let $r_p$ and $r_a$ be the spacecraft's geocentric radial distance at perigee and apogee, let ε be the eccentricity for the orbit or trajectory, let $\alpha_{eq}$ be the inclination to the equatorial plane, i.e., the directed angle from the z-axis of the $(x,y,z)$ frame to the Z-axis of the $(X,Y,Z)$ frame, and let $\theta$ be the parametric polar coordinate angle for the field point in the plane of the orbit or trajectory. Let $\theta$ range from $\theta_{min}$ to $\theta_{max}$. If the spacecraft is in a closed orbit, i.e., if 0≤ε<1, then $\theta_{min}$=−π and $\theta_{max}$=+π.

Let a spacecraft of negligible mass occupy the field point at $(r,\lambda,\phi)$, where $\lambda$ is the exterior geocentric latitude and $\phi$ is the exterior azimuthal angle. Let t be the observed coordinate time for the spacecraft. Let $r'$ be the retarded distance between an interior circulating point-mass dm and the exterior field point.

If an interior point-mass source emits a gravitational signal at $t'$, the signal will arrive at the field-point at a slightly later time t.

Assume the gravitational signal propagates at the speed $c_g$. Then the formula that connects t to t′ is

$$t = t' + \frac{r'}{c_g} \quad . \tag{2.1}$$

By definition, the Jacobian for the transformation from t to t′ is the derivative dt/dt′. Therefore, the Jacobian becomes

$$Jacobian = \frac{dt}{dt'} = 1 + \frac{1}{c_g}\frac{dr'}{dt'} \quad . \tag{2.2}$$

If dr′/dt′=0, then dt=dt′, in which case there is no effect of time retardation. This means that effects of time retardation are given by the Jacobian minus one.

$$Jacobian - 1 = \frac{dt}{dt'} - 1 = \frac{1}{c_g}\frac{dr'}{dt'} \quad . \tag{2.3}$$

The neoclassical causal version for Newton's law of gravity for a point-mass source is

$$\mathbf{g} = -\frac{GM_E}{r'^2}\frac{\mathbf{r'}}{r'} \quad , \tag{2.4}$$

where $\mathbf{r'}/r'$ is a unit vector directed towards increasing r′. Let $\mathbf{g}_r$ be the radial component of $\mathbf{g}$, and let $\mathbf{g}_{trt}$ be the time-retarded transverse component of $\mathbf{g}$. Then, the total gravitational field becomes

$$\mathbf{g} = \mathbf{g}_r + \mathbf{g}_{trt} \quad . \tag{2.5}$$

The component $\mathbf{g}_{trt}$ is perpendicular to $\mathbf{g}_r$.

$$\mathbf{g}_r \cdot \mathbf{g}_{trt} = 0 \quad . \tag{2.6}$$

The signed magnitude for the radial component is

$$g_r = -\frac{GM_E}{r^2} \quad . \tag{2.7}$$

The signed magnitude for the new causal time-retarded transverse component is

$$g_{trt} = -\frac{GM_E}{r'^2} TC_z(Jacobian-1) \quad , \tag{2.8}$$

where $TC_z$ is the relative magnitude for the Z-axis component of the transverse component.

The formulas for the radial distance for an elliptical orbit (0≤ε<1), for a parabolic trajectory (ε=1), or for a hyperbolic trajectory (ε>1), and its derivative, are

$$r(\theta) = \frac{r_p(1+\varepsilon)}{1+\varepsilon\cos(\theta)} \quad ,$$

$$\frac{dr}{d\theta} = \frac{r_p(1+\varepsilon)\varepsilon\sin(\theta)}{(1+\varepsilon\cos(\theta))^2} = \frac{r(\theta)^2}{r_p}\frac{\varepsilon}{1+\varepsilon}\sin(\theta) \quad . \tag{2.9}$$

Let $E$ be the spacecraft's orbital energy, the kinetic plus scalar potential energy, let $v(\theta)$ be the orbital speed, and let $m_{sc} \ll M_E$ be the mass of the spacecraft. A good first approximation for $E$ is given by conservation of orbital energy.

$$\text{constant} \cong E = \frac{1}{2} m_{sc} v(\theta)^2 - \frac{GM_E m_{sc}}{r(\theta)} = \frac{1}{2} m_{sc} v_p^2 - \frac{GM_E m_{sc}}{r_p} \quad , \tag{2.10}$$

where $v_p$ is the orbital speed at perigee. Solving for $v$ gives

$$v(\theta)^2 = v_p^2 - \frac{2GM_E}{r_p} + \frac{2GM_E}{r} \quad ,$$

$$v(\theta) = \left( v_p^2 - \frac{2GM_E}{r(\theta)} \left( \frac{r(\theta)}{r_p} - 1 \right) \right)^{\frac{1}{2}} \quad . \tag{2.11}$$

If the spacecraft is in a circular orbit, *i.e.*, if $\varepsilon=0$, then $r=r_p$, $dr/d\theta=0$, $v=v_p$, and the "geometrical acceleration" $v^2/r$ equals the radial gravitational acceleration $g_r$, $v^2/r=GM_E/r^2$.

Let $v_\perp$ be the component of $\mathbf{v}$ that is perpendicular to $\mathbf{r}$, and let $\Omega_\theta$ be the orbital angular speed, $\Omega_\theta \equiv d\theta/dt$. The formula for $v_\perp$ is

$$v_\perp(\theta) = r(\theta) \frac{d\theta}{dt} = r(\theta)\Omega_\theta(\theta) \quad . \tag{2.12}$$

Let $\mathbf{L}$ be the spacecraft's orbital angular momentum. The vector $\mathbf{L}$ is directed along the z-axis in the $(x,y,z)$ frame. Let $\Omega_p$ be the spacecraft's angular speed at perigee, and let $L_p$ be the magnitude of its angular momentum at perigee. By conservation of orbital angular momentum, the nearly constant absolute magnitude of $\mathbf{L}$ is

$$\text{constant} \cong L = m_{sc} v_\perp(\theta) r(\theta) = m_{sc} r(\theta)^2 \frac{d\theta}{dt}$$

$$= m_{sc} r(\theta)^2 \Omega_\theta(\theta) = L_p = m_{sc} r_p v_p = m_{sc} r_p^2 \Omega_p \quad . \tag{2.13}$$

To a good first approximation, the formula for the angular speed in the plane of the orbit or trajectory is

$$\Omega_\theta(\theta) \equiv \frac{d\theta}{dt} = \frac{r_p v_p}{r(\theta)^2} = \frac{r_p^2}{r(\theta)^2} \Omega_p \quad . \tag{2.14}$$

Let $\Delta t(\theta)$ be the time interval relative to perigee where $\theta=0$. The formulas for $\Delta t$ and the period $P$ are

$$\Delta t(\theta) = \int_0^\theta \frac{dt}{d\theta} d\theta = \int_0^\theta \frac{1}{\Omega_\theta(\theta)} d\theta \quad , \qquad P = \Delta t(\theta_{max}) - \Delta t(\theta_{min}) \quad . \tag{2.15}$$

Kepler's laws provide another method for finding the orbital angular speed and the period for closed orbits.[5] The 1st law states that the orbit is an ellipse, the 2nd law states that the area swept out by the orbital motion is proportional to the time interval for that motion, and the 3rd law states that the square of the period is proportional to the cube of the semimajor axis. Let $a$ and $b$ be the semimajor and semiminor axes. The 2$^{nd}$ law gives the orbital angular speed in terms of a, b, and P. The resulting formulas are as follows.

$$a = \frac{1}{2}\left(r_a + r_p\right) \ , \qquad b = a\left(1 - \varepsilon^2\right)^{\frac{1}{2}} \ ,$$

$$P = \frac{a^{\frac{3}{2}}}{\left(GM_E\right)^{\frac{1}{2}}} \ , \qquad \Omega_\theta(\theta) = \frac{2\pi}{P}\frac{ab}{r(\theta)^2} \ . \qquad (2.16)$$

Let $\theta_p$ be the parametric angle at which the orbit or trajectory intersects the equatorial plane. The X, Y, and Z-components of the geocentric radial distance to the field point, their time derivatives, and the components of **v** that are perpendicular to **r**, are

$$r_X = r\cos\left(\theta - \theta_p\right) \ ,$$

$$v_X = \frac{dr_X}{dt} = \left(\frac{dr}{d\theta}\cos\left(\theta - \theta_p\right)\right)\frac{d\theta}{dt} - \left(r\sin\left(\theta - \theta_p\right)\right)\frac{d\theta}{dt} \ ,$$

$$vX_\perp = -r\Omega_\theta\sin\left(\theta - \theta_p\right) \ ,$$

$$r_Y = r\cos\left(\alpha_{eq}\right)\sin\left(\theta - \theta_p\right) \ ,$$

$$v_Y = \frac{dr_Y}{dt} = \left(\frac{dr}{d\theta}\cos\left(\alpha_{eq}\right)\sin\left(\theta - \theta_p\right)\right)\frac{d\theta}{dt} + \left(r\cos\left(\alpha_{eq}\right)\cos\left(\theta - \theta_p\right)\right)\frac{d\theta}{dt} \ ,$$

$$vY_\perp = r\Omega_\theta\cos\left(\alpha_{eq}\right)\cos\left(\theta - \theta_p\right) \ ,$$

$$r_Z = -r\sin\left(\alpha_{eq}\right)\sin\left(\theta - \theta_p\right) \ ,$$

$$v_Z = \frac{dr_Z}{dt} = -\left(\frac{dr}{d\theta}\sin\left(\alpha_{eq}\right)\sin\left(\theta - \theta_p\right)\right)\frac{d\theta}{dt} - \left(r\sin\left(\alpha_{eq}\right)\cos\left(\theta - \theta_p\right)\right)\frac{d\theta}{dt} \ ,$$

$$vZ_\perp = -r\Omega_\theta\sin\left(\alpha_{eq}\right)\cos\left(\theta - \theta_p\right) \ . \qquad (2.17)$$

Perigee is at $\theta=0$, and the orbit intersects the equatorial plane at $\theta=\theta_p$.

The value for the angle $\theta_p$ depends on the latitude for perigee $\lambda_p$, which ranges from -90° to +90°, and the inclination $\alpha_{eq}$, which ranges from 0° to 180°. If $\alpha_{eq}$=0° or 180°, then $\theta_p$=0°. If $\alpha_{eq}\neq0$° or 180° and $\sin\left(\lambda_p\right)\leq\sin\left(\alpha_{eq}\right)$, the formula for $\theta_p$ (in radians) is

$$\theta_p = \sin^{-1}\left(\frac{\sin\left(\lambda_p\right)}{\sin\left(\alpha_{eq}\right)}\right) \ . \qquad (2.18)$$

If $\sin\left(\lambda_p\right)>\sin\left(\alpha_{eq}\right)$, the inverse sine function is shifted from the primary branch and the value for $\theta_p$ is greater than 90°.

Let $\phi$ be the azimuthal angle for the projection of the field point onto the (X,Y) equatorial plane, and let $\Omega_\phi$ be the angular speed for $\phi$. Then

$$\Omega_\phi \equiv \frac{d\phi}{dt} = \frac{d\phi}{d\theta}\frac{d\theta}{dt} = \Omega_\theta\frac{d\phi}{d\theta} \ , \qquad \frac{d\phi}{d\theta} = \frac{\Omega_\phi}{\Omega_\theta} \ . \qquad (2.19)$$

Let $r_\phi$ be the geocentric radial distance to the projection of the field point onto the (X,Y) equatorial plane, and let $r_\lambda$ be the geocentric radial distance to the projection of the field point onto the (X,Z) plane. Then

$$r_\phi(\theta) = \left(r_X^2 + r_Y^2\right)^{\frac{1}{2}} = r(\theta)\left(\cos^2(\theta - \theta_p) + \cos^2(\alpha_{eq})\sin^2(\theta - \theta_p)\right)^{\frac{1}{2}} \ ,$$

$$r_\lambda(\theta) = \left(r_X^2 + r_Z^2\right)^{\frac{1}{2}} = r(\theta)\left(\cos^2(\theta - \theta_p) + \sin^2(\alpha_{eq})\sin^2(\theta - \theta_p)\right)^{\frac{1}{2}} \ . \tag{2.20}$$

The formula for the geocentric latitude, $\lambda$, is

$$\lambda(\theta) = \tan^{-1}\left(\frac{r_Z}{r_\phi}\right) = \tan^{-1}\left(\frac{\sin(\alpha_{eq})\sin(\theta - \theta_p)}{\left(\cos^2(\theta - \theta_p) + \cos^2(\alpha_{eq})\sin^2(\theta - \theta_p)\right)^{\frac{1}{2}}}\right) \ . \tag{2.21}$$

Let $v\phi_\perp$ be the component of $\mathbf{v}_\phi$ that is perpendicular to $\mathbf{r}_\phi$, such that

$$v\phi_\perp = r_\phi\frac{d\phi}{dt} = r_\phi\Omega_\phi \ . \tag{2.22}$$

The formula for $v\phi_\perp$ is also given by the perpendicular components of $\mathbf{v}$ from Eq (2.17).

$$v\phi_\perp = \left(vX_\perp^2 + vY_\perp^2\right)^{\frac{1}{2}} = r\Omega_\theta\left(\sin^2(\theta - \theta_p) + \cos^2(\alpha_{eq})\cos^2(\theta - \theta_p)\right)^{\frac{1}{2}} \ .$$

Therefore,

$$\Omega_\phi = \frac{v\phi_\perp}{r_\phi} = \Omega_\theta\left(\frac{\sin^2(\theta - \theta_p) + \cos^2(\alpha_{eq})\cos^2(\theta - \theta_p)}{\cos^2(\theta - \theta_p) + \cos^2(\alpha_{eq})\sin^2(\theta - \theta_p)}\right)^{\frac{1}{2}} \ ,$$

$$= \Omega_\theta\left(\frac{\cos^2(\alpha_{eq}) + \tan^2(\theta - \theta_p)}{1 + \cos^2(\alpha_{eq})\tan^2(\theta - \theta_p)}\right)^{\frac{1}{2}} \ . \tag{2.23}$$

The formula for $d\phi/d\theta$ is

$$\frac{d\phi}{d\theta} = \frac{\Omega_\phi}{\Omega_\theta} = \left(\frac{\cos^2(\alpha_{eq}) + \tan^2(\theta - \theta_p)}{1 + \cos^2(\alpha_{eq})\tan^2(\theta - \theta_p)}\right)^{\frac{1}{2}} \ , \tag{2.24}$$

and the formula for $\phi(\theta)$ becomes

$$\phi(\theta) = \int_0^\theta\left(\frac{\cos^2(\alpha_{eq}) + \tan^2(\theta - \theta_p)}{1 + \cos^2(\alpha_{eq})\tan^2(\theta - \theta_p)}\right)^{\frac{1}{2}} d\theta \ . \tag{2.25}$$

The components of the radial distance to the spacecraft can now be rewritten in terms of the orthogonal variables r, $\lambda$, and $\phi$.

$$r_X = r\cos(\lambda)\cos(\phi) \ ,$$
$$r_Y = r\cos(\lambda)\sin(\phi) \ ,$$
$$r_Z = r\sin(\lambda) \ . \tag{2.26}$$

A reasonably valid formula for the Earth's radial mass-density distribution can be found in ref. [2].

$$\rho(r) = \text{if}\left(r < r_{ic}, \rho_{ic}, \text{if}\left(r < r_{oc}, \rho_{oc}(r), \text{if}\left(r < r_{man}, \rho_{man}(r), \rho_{cst}(r)\right)\right)\right) \ , \tag{2.27}$$

where $r_{ic}$ is the radius for the inner core, $r_{oc}$ is the outer radius for the outer core, $r_{man}$ is that for the mantle, and $r_E$ is the Earth's spherical radius. Numerical values and formulas for $r_{ic}$, $\rho_{ic}$, $r_{oc}$, $\rho_{oc}$, *etc*, can be found in ref. [2].

Divide the sphere into a sequence of elemental point-mass sources, dm. The formula for dm is

$$dm = \rho(r)r^2 \cos(\lambda')dr d\lambda' d\phi' \ . \tag{2.28}$$

The X, Y, and Z components of the radial vector for the source-point, **r**, are

$$r_X = r \cos(\lambda') \cos(\phi') \ ,$$
$$r_Y = r \cos(\lambda') \sin(\phi') \ ,$$
$$r_Z = r \sin(\lambda') \ . \tag{2.29}$$

The retarded distance r' is the distance from a circulating interior point-mass dm at **r** to the exterior field point at **r**. The X, Y, and Z-components of the retarded distance are

$$r'_X = r_X - r_X = r \cos(\lambda) \cos(\phi) - r \cos(\lambda') \cos(\phi') \ ,$$
$$r'_Y = r_Y - r_Y = r \cos(\lambda) \sin(\phi) - r \cos(\lambda') \sin(\phi') \ ,$$
$$r'_Z = r_Z - r_Z = r \sin(\lambda) - r \sin(\lambda') \ . \tag{2.30}$$

The magnitude r' is

$$r' = \left(r'^2_x + r'^2_y + r'^e_z\right)^{\frac{1}{2}}$$
$$= \left(r^2 + r^2 - 2rr\left(\cos(\lambda)\cos(\lambda')\cos(\phi - \phi') + \sin(\lambda)\sin(\lambda')\right)\right)^{\frac{1}{2}} \ ,$$

which can be rewritten as

$$r' = r\left(1 + x\right)^{\frac{1}{2}} \ , \tag{2.31}$$

where x is defined by

$$x \equiv \frac{r^2}{r^2} - 2\frac{r}{r}\left(\cos(\lambda)\cos(\lambda')\cos(\phi - \phi') + \sin(\lambda)\sin(\lambda')\right) \ . \tag{2.32}$$

The retarded-time derivative of r' is

$$\frac{dr'}{dt'} = \frac{\partial r'}{\partial \phi}\frac{d\phi}{dt} + \frac{\partial r'}{\partial \phi'}\frac{d\phi'}{dt'} = \Omega_\phi\frac{\partial r'}{\partial \phi} + \Omega_E\frac{\partial r'}{\partial \phi'} \ ,$$

and reduces to

$$\frac{dr'}{dt'} = \frac{rr}{r'}\left(\Omega_\phi - \Omega_E\right)\cos(\lambda)\cos(\lambda')\sin(\phi - \phi') \ . \tag{2.33}$$

The Jacobian-1 becomes

$$\frac{1}{c_g}\frac{dr'}{dt'} = \frac{r_E\Omega_E}{c_g}\frac{r}{r'}\frac{r}{r_E}\frac{\Omega_\phi - \Omega_E}{\Omega_E}\cos(\lambda)\cos(\lambda')\sin(\phi - \phi') \quad . \tag{2.34}$$

The relative transverse component **TC** of the vector **r′** is given by the vector cross product, **r×r′**/rr′, such that

$$\mathbf{TC} = \frac{\mathbf{r} \times \mathbf{r'}}{rr'} = \frac{\mathbf{r} \times (\mathbf{r} - \mathbf{r})}{rr'} = \frac{\mathbf{r} \times \mathbf{r} - \mathbf{r} \times \mathbf{r}}{rr'} = \frac{0 - \mathbf{r} \times \mathbf{r}}{rr'} = \frac{\mathbf{r} \times \mathbf{r}}{rr'} \quad .$$

Written in component form,

$$\mathbf{TC} = \frac{1}{rr'}\left(\left(r_y r_z - r_y r_z\right)_X, \left(r_z r_x - r_z r_x\right)_Y, \left(r_x r_y - r_x r_y\right)_Z\right) \quad . \tag{2.35}$$

The magnitude for the Z-component of **TC** is

$$TC_Z = \frac{1}{rr'}\left(r_x r_y - r_x r_y\right) = \frac{r}{r'}\cos(\lambda)\cos(\lambda')\sin(\phi - \phi') \quad . \tag{2.36}$$

Let $d^3g_{trt}$ be the differential form for the magnitude of the time-retarded transverse component of the gravitational field. The starting formula for $d^3g_{trt}$ is

$$d^3g_{trt} = \left(\frac{1}{r'^2}\text{ gravity law}\right)(TC_Z)(\text{Jacobian-1}) \quad .$$

Substituting Eqs. (2.28), (2.34), and (2.36) into the starting formula gives

$$d^3g_{trt} = \left(-G\frac{\rho(r)r^2\cos(\lambda')dr d\lambda' d\phi'}{r'^2}\right)\left(\frac{r}{r'}\cos(\lambda)\cos(\lambda')\sin(\phi - \phi')\right)$$

$$\times \left(\frac{r_E\Omega_E}{c_g}\frac{r}{r'}\frac{r}{r_E}\frac{\Omega_\phi - \Omega_E}{\Omega_E}\cos(\lambda)\cos(\lambda')\sin(\phi - \phi')\right) \quad .$$

By rearrangement, $d^3g_{trt}$ can be rewritten as the product of a coefficient, A, the ratio of the angular speeds, and the integrand, IG, which is

$$d^3g_{trt} = -A\left(\frac{\Omega_\phi - \Omega_E}{\Omega_E}\right)IG\,\frac{dr}{r_E}d\lambda' d\phi' \quad , \tag{2.37}$$

where the definitions for the equatorial surface speed, $v_{Eq}$, and for the coefficient A, are

$$v_{Eq} \equiv r_E\Omega_E \quad , \qquad A \equiv \left(\frac{v_{Eq}}{c_g}\right)\left(G\bar{\rho}_E r_E\right) \quad , \tag{2.38}$$

and the integrand for the triple integration is defined by

$$IG \equiv \cos^2(\lambda)\left(\frac{\rho(r)}{\bar{\rho}_E}\frac{r^4}{r_E^4}\right)\left(\cos^3(\lambda')\right)\left(\frac{r_E}{r}\right)^3\left(\frac{\sin^2(\phi - \phi')}{(1 + x)^2}\right) \quad , \tag{2.39}$$

where $\bar{\rho}_E$ is the mean value for $\rho_E(r)$.

Let $Ig\phi'$ be the integral over $\phi'$, let $Ig\lambda'$ be the integral over $\lambda'$, let $Igr$ be the integral over $r$, and let $Ig$ be the product $\cos^2(\lambda) \times Igr$, defined as follows.

$$Ig\phi' \equiv \left(\frac{r_E}{r}\right)^3 \int_{-\pi}^{\pi} \frac{\sin^2(\phi - \phi')}{(1 + x)^2} d\phi' \quad ,$$

$$Ig\lambda' \equiv \int_{-\pi/2}^{\pi/2} Ig\phi' \cos^3(\lambda') d\lambda' \quad ,$$

$$Igr \equiv \int_0^{r_E} Ig\lambda' \frac{\rho(r)}{\bar{\rho}_E} \frac{r^4}{r_E^4} \frac{dr}{r_E} \quad ,$$

$$Ig \equiv \cos^2(\lambda) Igr \quad . \tag{2.40}$$

The triple integral, $Igr$ , can be solved by using numerical integration, but it takes a lot of computer time, particularly if $r$ is near $r_E$ (the singularity at $r=r_E$ must be avoided). To speed up the calculation, an algebraic expression for $Igr$ is needed.

By using numerical integration, it can be shown that $Igr$ is independent of $\lambda$ and $\phi$, which means that $Igr$ can be computed with $\lambda=0$ and $\phi=0$.

Assume that $Igr(r)$ can be approximated by a power series. Let $PSr(r)$ be a four-term power series, defined as follows.

$$PSr(r) \equiv \left(\frac{I_E}{\bar{\rho}_E r_E^5}\right)\left(\frac{r_E}{r}\right)^3 \left(C_0 + C_2\left(\frac{r_E}{r}\right)^2 + C_4\left(\frac{r_E}{r}\right)^4 + C_6\left(\frac{r_E}{r}\right)^6\right) \quad . \tag{2.41}$$

The coefficients can be adjusted to give an optimum fit to values for $Igr$ calculated by using numerical integration. By using a least-squares fitting routine, the following values for the coefficients were found to give an excellent fit of $PSr(r)$ to $Igr(r)$.

$C_0 = 0.50889$ , $\quad C_2 = 0.13931$ ,

$C_4 = 0.01013$ , $\quad C_6 = 0.14671$ . $\tag{2.42}$

The maximum difference between $Igr$ and $PSr$ is less than $1 \times 10^{-5}$. Notice that $PSr(r)$ can be extrapolated all the way down to the surface where $r=r_E$.

Let $PS(r)$ be the same power series without the coefficient, defined as follows.

$$PS(r) \equiv \left(\frac{r_E}{r}\right)^3 \left(C_0 + C_2\left(\frac{r_E}{r}\right)^2 + C_4\left(\frac{r_E}{r}\right)^4 + C_6\left(\frac{r_E}{r}\right)^6\right) \quad . \tag{2.43}$$

The solution for $g_{trt}(\theta)$ can now be written as

$$g_{trt}(\theta) = -A_{trt}\left(\frac{\Omega_\phi(\theta) - \Omega_E}{\Omega_E}\right)\cos^2(\lambda(\theta))PS(r(\theta)) \quad , \tag{2.44}$$

where the definition for the new constant $A_{trt}$ along with its numerical value (with $c_g = c$) is

$$A_{trt} \equiv G \frac{I_E}{r_E^4} \frac{v_{Eq}}{c_g} = 5.0364 \times 10^{-6} \frac{m}{s^2} \quad . \tag{2.45}$$

## 3. Time-retarded transverse induction-like field of a central spinning sphere

Let $\mathbf{e}_r$, $\mathbf{e}_\lambda$, and $\mathbf{e}_\phi$ be orthogonal unit vectors for the spherical system $(r,\lambda,\phi)$; $\mathbf{e}_r$ is directed radially outward, $\mathbf{e}_\lambda$ is directed southward, and $\mathbf{e}_\phi$ is directed eastward. The polar angle or colatitude increases towards $+\mathbf{e}_\lambda$ (southward); the latitude $\lambda$ increases towards $-\mathbf{e}_\lambda$ (northward).

The vector velocity for the spacecraft in the $(r,\lambda,\phi)$ system becomes

$$\mathbf{v} = \mathbf{e}_r v_r + \mathbf{e}_\lambda v_\lambda + \mathbf{e}_\phi v_\phi \quad . \tag{3.1}$$

The $r$, $\lambda$, and $\phi$-components of $\mathbf{v}$ are as follows.

$$v_r = \frac{dr}{dt} = \Omega_0 \frac{dr}{d\theta} \quad ,$$

$$v_\lambda = r_\lambda \frac{d\lambda}{dt} = r_\lambda \Omega_0 \frac{d\lambda}{d\theta} \quad ,$$

$$v_\phi = r_\phi \frac{d\phi}{dt} = r_\phi \Omega_0 \frac{d\phi}{d\theta} \quad . \tag{3.2}$$

The transverse acceleration field, $\mathbf{g}_{trt} = -\mathbf{e}_\phi g_{trt}$, is a solenoidal or vortex field.[6] The divergence of $\mathbf{g}_{trt}$ is zero, which means it cannot be derived from the gradient of a scalar potential. But the curl of $\mathbf{g}_{trt}$ is not zero, which means it can be derived from the curl of a vector potential. The empirical prediction formula (*cf.* Eq. (1.1)) indicates that the speed change must be due to a change in the component of $\mathbf{v}$ that is directed along $\mathbf{e}_\lambda$. But the transverse field $\mathbf{g}_{trt}$, being directed along $\mathbf{e}_\phi$, cannot directly change the <u>magnitude</u> of $\mathbf{v}_\lambda$. However, a time rate of change of $\mathbf{g}_{trt}$ can, by mimicry of induction fields, generate a vector field that is transverse to $\mathbf{g}_{trt}$ and is directed along $\mathbf{e}_\lambda$.

Let $\mathbf{F}_\lambda$ be an induction-like field that is directed along $\mathbf{e}_\lambda$. Assume that the $\mathbf{e}_\phi$ component of the curl of $\mathbf{F}_\lambda$ equals $-kd\mathbf{g}_e/dt$, where $k$ is a constant of proportionality. The formula for the curl operation in spherical coordinates can be found in J. D. Jackson's textbook.[7]

$$\nabla \times \mathbf{F}_\lambda = \mathbf{e}_\phi \frac{1}{r} \frac{\partial}{\partial r} (rF_\lambda) = -k \frac{d\mathbf{g}_{trt}}{dt} = \mathbf{e}_\phi k \frac{dg_{trt}}{dt} \quad . \tag{3.3}$$

Solving for $\partial(rF_\lambda)/\partial r$ and integrating both sides from $t(0)$ to $t(\theta)$ gives

$$\int_{t(0)}^{t(\theta)} \frac{\partial}{\partial r} (rF_\lambda) \, dt = \int_0^\theta \frac{\partial}{\partial r} (rF_\lambda) \frac{dt}{d\theta} \, d\theta = \int_0^\theta \frac{d}{d\theta} (rF_\lambda) \frac{d\theta}{dr} \frac{dt}{d\theta} \, d\theta$$

$$= k \int_{t(0)}^{t(\theta)} r \frac{dg_{trt}}{dt} \, dt = k \int_0^\theta r \frac{dg_{trt}}{d\theta} \, d\theta \quad . \tag{3.4}$$

Therefore,

$$\int_0^\theta \left( \frac{d}{d\theta}\left(rF_\lambda\right)\frac{d\theta}{dr}\frac{dt}{d\theta} - kr\frac{dg_{trt}}{d\theta} \right) d\theta = 0 \quad . \tag{3.5}$$

This equation is satisfied for all values of $\theta$, if and only if,

$$\frac{d}{d\theta}\left(rF_\lambda\right) = kr\frac{dr}{d\theta}\frac{d\theta}{dt}\frac{dg_{trt}}{d\theta} \quad , \tag{3.6}$$

so that,

$$F_\lambda(\theta) = \frac{k}{r(\theta)}\int_0^\theta r(\theta)\Omega_0(\theta)\frac{dr}{d\theta}\frac{dg_{trt}}{d\theta}\,d\theta \quad . \tag{3.7}$$

Units for $F_\lambda$ are m/s$^2$, i.e., $F_\lambda$ has the same units as those for $g_r$ and $g_{trt}$. The constant k has units of $(m/s)^{-1}$. Let $v_k$ be the reciprocal of k, $v_k\equiv1/k$. Regard $v_k$ as an adjustable parameter that will be called the "induction speed". Regard the Earth's equatorial surface speed, $v_{Eq}=r_E\Omega_E$, as a convenient reference speed.

The formula for $F_\lambda$ can be rewritten in terms of $v_k$ and $v_{Eq}$.

$$F_\lambda(\theta) = \frac{v_{Eq}}{v_k}\frac{r_E}{r(\theta)}\int_0^\theta \frac{r(\theta)}{r_E}\frac{\Omega_0(\theta)}{\Omega_E}\frac{1}{r_E}\frac{dr}{d\theta}\frac{dg_E}{d\theta}\,d\theta \quad . \tag{3.8}$$

## 4. Speed-change caused by the induction-like field of a central spinning sphere

The induction-like field $\mathbf{F}_\lambda$ is an acceleration field that is directed along $\pm\mathbf{e}_\lambda$, which means that $\mathbf{F}_\lambda$ can change the magnitude of the $\lambda$-component of the velocity, $\mathbf{v}_\lambda$.

The dot product $\mathbf{v}\bullet\mathbf{F}_\lambda$ gives the rate of change in the square of $v_\lambda$, i.e., the "anomalous" rate of change in the spacecraft's kinetic energy. This dot product gives

$$\mathbf{v}\bullet\mathbf{F}_\lambda = \left(\mathbf{e}_r v_r + \mathbf{e}_\lambda v_\lambda + \mathbf{e}_\phi v_\phi\right)\bullet\mathbf{e}_\lambda F_\lambda = v_\lambda F_\lambda = r_\lambda\Omega_0 F_\lambda \frac{d\lambda}{d\theta} \quad , \tag{4.1}$$

where the formula for $r_\lambda$ is given by Eq. (2.20).

Let $\delta v/v_{in}$ be the relative change in the magnitude of $\mathbf{v}_\lambda$ due to $\mathbf{F}_\lambda$, where $v_{in}$ is the initial speed, such that

$$\left(1 + \frac{\delta v}{v_{in}}\right)^2 \cong 1 + 2\frac{\delta v}{v_{in}} = 1 + \frac{1}{v_{in}^2}\int_{t(0)}^{t(\theta)} F_\lambda v_\lambda dt = 1 + \frac{1}{v_{in}^2}\int_0^\theta r_\lambda F_\lambda \frac{d\lambda}{d\theta}\,d\theta \quad . \tag{4.2}$$

Therefore,

$$\delta v(\theta) = \frac{v_{in}}{2}\int_0^\theta \frac{r_\lambda(\theta)F_\lambda(\theta)}{v_{in}^2}\frac{d\lambda}{d\theta}\,d\theta \quad , \tag{4.3}$$

and

$$\delta v_{in} = \delta v(\theta_{min}) = \frac{v_{in}}{2} \int_0^{\theta_{min}} \frac{r_\lambda(\theta) F_\lambda(\theta)}{v_{in}^2} \frac{d\lambda}{d\theta} d\theta \ ,$$

$$\delta v_{out} = \delta v(\theta_{max}) = \frac{v_{in}}{2} \int_0^{\theta_{max}} \frac{r_\lambda(\theta) F_\lambda(\theta)}{v_{in}^2} \frac{d\lambda}{d\theta} d\theta \ . \tag{4.4}$$

Let $\delta v_{trt}$ be the total speed change for the neoclassical causal version of Newton's theory.

$$\delta v_{trt} = \delta v_{in} + \delta v_{out} \ . \tag{4.5}$$

These are the formulas that were used in ref. [2] to calculate the speed change listed in Table I for each of the six flybys reported by Anderson *et al*.[1]

## 5. Anomalous decrease in the radius of the Moon's orbit

For a reasonably valid first approximation, regard the Earth and Moon as an isolated binary system. Let $(x,y,z)$ be a nonrotating barycentric frame of reference, and let the $(x,y)$ plane coincide with the Earth-Moon orbital plane. The center-of-mass coincides with the origin for this reference frame.

Let the Moon be simulated by a point mass of mass $M_M$. Numerical values for the Moon are listed in Appendix B.

Numerical values for $M_E$, $M_M$, the reduced mass, and the ratios $M_E/(M_E+M_M)$, $M_M/(M_E+M_M)$, and $M_E/M_M$, are

$$M_E = 5.9761 \times 10^{24} \text{ kg} \ , \qquad \frac{M_E}{M_E + M_M} = 0.9879 \ ,$$

$$M_M = 7.3477 \times 10^{22} \text{ kg} \ , \qquad \frac{M_M}{M_E + M_M} = 0.0121 \ ,$$

$$\frac{M_E M_M}{M_E + M_M} = 7.2585 \times 10^{22} \text{ kg} \qquad \frac{M_E}{M_M} = 81.3329 \ . \tag{5.1}$$

The Moon and Earth revolve around the center-of-mass in similar confocal elliptical orbits. Let $r(\theta)$ be the radial distance from the center of the Earth to the center of the Moon, and let $r_a$ and $r_p$ be the value for $r$ at apogee ($\theta=\pm\pi$) and at perigee ($\theta=0$). Let $\varepsilon$ be the eccentricity for the ellipses. Then

$$r(\theta) = \frac{r_p(1 + \varepsilon)}{1 + \varepsilon \cos(\theta)} \ , \qquad\qquad \frac{dr}{d\theta} = \frac{r(\theta)^2}{r_p} \frac{\varepsilon}{1 + \varepsilon} \sin(\theta) \ ,$$

$$\varepsilon = \frac{r_a - r_p}{r_a + r_p} = 0.0554 \ ,$$

$$r_a = r(\pm\pi) = 405696 \text{ km} = 63.6782 \ r_E \ ,$$

$$r_p = r(0) = 363104 \text{ km} = 56.9929 \ r_E \ . \tag{5.2}$$

Let $rM(\theta)$ be the radial distance from the origin to the Moon, and let $rM_a$ and $rM_p$ be the value for $rM$ at apogee and at perigee. Let $a_M$ and $b_M$ be the semimajor and semiminor axes for the Moon's elliptical orbit.

Then

$$rM(\theta) = \frac{M_E}{M_E + M_M} r(\theta) = 0.9879 \, r(\theta) \quad , \qquad \frac{drM}{d\theta} = \frac{M_E}{M_E + M_M} \frac{dr}{d\theta} \quad ,$$

$$rM_a = \frac{M_E}{M_E + M_M} r_a = 62.9048 \, r_E \quad ,$$

$$rM_p = \frac{M_E}{M_E + M_M} r_p = 56.3007 \, r_E \quad ,$$

$$a_M = \frac{1}{2} \left( rM_a + rM_p \right) = 59.6028 \, r_E \quad ,$$

$$b_M = aM \left( 1 - \varepsilon^2 \right)^{\frac{1}{2}} = 59.5112 \, r_E \quad .$$

(5.3)

The lunar period can be calculated by using Kepler's 3$^{rd}$ Law.[5] Let $P_M$ be the calculated period for the isolated Earth-Moon binary system.

$$P_M = \frac{2\pi}{\left( G \left( M_E + M_M \right) \right)^{\frac{1}{2}}} a_M^{\frac{3}{2}} = 2.3140 \times 10^6 s = 26.7825 \text{ days} \quad .$$

(5.4)

Let $v_{co}$ be the orbital speed for an equivalent circular orbit which has the radius $a_M$ and the period $P_M$. The above formula for $P_M$ can be rewritten to display the formula for $v_{co}$.

$$P_M = \frac{2\pi a_M}{\left( G \left( M_E + M_M \right) \right)^{\frac{1}{2}}} a_M^{\frac{1}{2}} = \frac{2\pi a_M}{\left( G \left( M_E + M_M \right)/a_M \right)^{\frac{1}{2}}} = \frac{2\pi a_M}{v_{co}} \quad ,$$

$$v_{co} = \left( \frac{G \left( M_E + M_M \right)}{a_M} \right)^{\frac{1}{2}} = 1031.0779 \text{ m/s} \quad .$$

(5.5)

Let $\delta v_{co} \ll v_{co}$ be a small change in the orbital speed. Assume that the cause for $\delta v_{co}$ is a transverse perturbation which does not change the radial gravitational field. In this case, the geometrical acceleration, $v_{co}^2/a_M$, which equals the radial gravitational acceleration, is constant. Let $\delta a_M \ll a_M$ be the corresponding change in the radius of the orbit. Then

$$\text{constant} \cong 1 = \frac{\left( 1 + \delta v_{co}/v_{co} \right)^2}{\left( 1 + \delta a_M/a_M \right)} \cong 1 + 2 \frac{\delta v_{co}}{v_{co}} - \frac{\delta a_M}{a_M} \quad ,$$

$$2 \frac{\delta v_{co}}{v_{co}} \cong \frac{\delta a_M}{a_M} \quad .$$

(5.6)

There are two cases to consider: (1) according to Stephenson and Morrison,[3] tidal braking increases the LOD by $+23 \times 10^{-6}$ s per year, and (2) lunar laser ranging experiments have shown that the radius of the Moon's orbit is increasing at the rate of +38 mm per year.[4]

Case 1. Let day$_{sol}$ be the current number of seconds in a solar day (by the Sun), let day$_{sid}$ be the current number of seconds in a sidereal day

(by the stars), and let K be the ratio $day_{sol}/day_{sid}$. Then

$$day_{sol} = 60 \times 60 \times 24 = 86400 \, s \quad,$$

$$day_{sid} = \frac{2\pi}{\Omega_E} = 86164.1 \, s \quad,$$

$$K = \frac{day_{sol}}{day_{sid}} = 1.002738 \quad,$$

$$LOD = \frac{2\pi K}{\Omega_E} = 86400 \, s \quad. \tag{5.7}$$

Let $\delta LOD$ be the rate of change in the LOD caused by tidal braking. Then

$$\delta LOD = 23 \times 10^{-6} \, s \text{ per year} \quad. \tag{5.8}$$

Let $\delta\Omega_E$ be a small change in $\Omega_E$. Then

$$1 + \frac{\delta LOD}{LOD} = \frac{1}{\left(1 + \delta\Omega_E/\Omega_E\right)} \cong 1 - \frac{\delta\Omega_E}{\Omega_E} \quad,$$

$$\delta\Omega_E = -\Omega_E \frac{\delta LOD}{LOD} = -1.9412 \times 10^{-14} \, rad/s \text{ per year} \quad. \tag{5.9}$$

Let $S_E$ be the Earth's spin angular momentum, and let $\delta S_E$ be a small change in $S_E$. Assume there is no change in $I_E$. Then

$$S_E = I_E\Omega_E = 5.8510 \times 10^{33} \, kg{\cdot}m^2/s \quad,$$

$$\delta S_E = S_E \frac{\delta\Omega_E}{\Omega_E} = -1.5576 \times 10^{24} \, kg{\cdot}m^2/s \text{ per year} \quad. \tag{5.10}$$

Let $L_M$ be the magnitude for the Moon's orbital angular momentum, and let $\delta L_M$ be a small change in $L_M$. Then, by conservation of angular momentum,

$$L_M = M_M v_{co} a_M = 2.8769 \times 10^{34} \, kg{\cdot}m^2/s \quad,$$

$$L_M + S_E = constant \quad,$$

$$\delta L_M = -\delta S_E = +1.5576 \times 10^{24} \, kg{\cdot}m^2/s \text{ per year} \quad. \tag{5.11}$$

Let $\delta v_{co}$ be a small change in the Moon's orbital speed. Then (*cf* Eq. (5.6))

$$\frac{\delta L_M}{L_M} = \left(\frac{\delta v_{co}}{v_{co}} + \frac{\delta a_M}{a_M}\right) = \left(3 \frac{\delta v_{co}}{v_{co}}\right) \quad,$$

$$\delta v_{co} = \frac{v_{co}}{3} \frac{\delta L_M}{L_M} = +18.6 \times 10^{-9} \, m/s \text{ per year} \quad. \tag{5.12}$$

Solving Eq. 5.6 for $\delta a_M$ gives

$$\delta a_M = 2a_M \frac{\delta v_{co}}{v_{co}} = +13.7 \times 10^{-3} \, m \text{ per year} \quad. \tag{5.13}$$

Thus we find that tidal braking alone causes an increase in the radius for the Moon's orbit of 14 mm per year, and a corresponding increase in the Moon's orbital speed of $19 \times 10^{-9}$ m/s per year.

Case 2. For this case,

$$\delta a_M = +38 \times 10^{-3} \text{ m per year} \quad . \tag{5.14}$$

Substituting this value for $\delta a_M$ into Eq. (5.6), the value from Eq. (5.3) for $a_M$, and the value from Eq. (5.5) for $v_{co}$, gives the following rate for the observed increase in the orbital speed.

$$\delta v_{co} = \frac{v_{co}}{2} \frac{\delta a_M}{a_M} = +51.6 \times 10^{-9} \text{ m/s per year} \quad . \tag{5.15}$$

There is an obvious difference between Case 1 and Case 2. An unexplained action is causing the rate for change in the radius to decrease from 38 to 14 mm per year (-24 mm per year), and for the corresponding rate for change in the orbital speed to decrease from 52 to $19 \times 10^{-9}$ m/s per year ($-33 \times 10^{-9}$ m/s per year). This unexplained difference is the "lunar orbit anomaly".

## 6. Orbital speed-change caused by the new causal version of Newton's theory

The radial distance from the center-of-mass to the Moon, rM($\theta$), and its derivative, drM/d$\theta$, are given by Eqs. (5.2) and (5.3).

The Moon's orbital angular speed is given by Eq. (2.16).

$$\Omega_\theta(\theta) = \frac{2\pi}{P_M} \frac{a_M b_M}{rM(\theta)^2} \quad , \tag{6.2}$$

where $a_M$, and $b_M$ are given by Eq. (5.3) and $P_M$ is given by Eq. (5.4).

Let $\Omega_p$ and $\Omega_a$ be the angular speed at perigee and apogee.

$$\Omega_p = \Omega_\theta(0) = 3.0385 \times 10^{-6} \text{ rad/s} \quad , \qquad \frac{\Omega_E}{\Omega_p} = 23.9994 \quad ,$$

$$\Omega_a = \Omega_\theta(\pm\pi) = 2.4340 \times 10^{-6} \text{ rad/s} \quad , \qquad \frac{\Omega_E}{\Omega_a} = 29.9599 \quad . \tag{6.3}$$

The formula for the Moon's orbital speed is

$$vM(\theta) = rM(\theta) \left( 1 + \left( \frac{1}{rM(\theta)} \frac{drM}{d\theta} \right)^2 \right)^{\frac{1}{2}} \Omega_\theta(\theta) \quad . \tag{6.4}$$

Let $vM_p$ and $vM_a$ be the Moon's orbital speed at perigee and at apogee. Compare $vM_p$ and $vM_a$ with $v_{co}$ (cf Eq. (5.5)).

$$vM_p = \Omega_p rM_p = 1089.8740 \text{ m/s} \quad ,$$

$$vM_a = \Omega_a rM_a = 975.4536 \text{ m/s} \quad ,$$

$$v_{co} = 1031.0779 \text{ m/s} \quad . \tag{6.5}$$

The average inclination for the Moon's orbital plane is (from Appendix B)

$$\alpha_{Eq} = 23.43° \pm 5.15° \quad . \tag{6.6}$$

If the latitude for perigee, $\lambda_p = \alpha_{Eq}$, and the parametric angle $\theta = 0$ at perigee, then the formula for geocentric latitude for the Moon becomes

$$\lambda(\theta) = \tan^{-1}\left(\tan(\alpha_{Eq})\cos(\theta)\right) \quad . \tag{6.7}$$

The $\phi$-component of the angular speed reduces to

$$\Omega_\phi(\theta) = \Omega_0(\theta)\cos(\alpha_{Eq}) = 0.9175\,\Omega_0(\theta) \quad . \tag{6.8}$$

The formula for the $\lambda$-component of rM becomes

$$r_\lambda(\theta) = rM(\theta)\cos(\theta) \quad . \tag{6.9}$$

The formula for $g_{trt}$ is given by Eq. (2.44). If $c_g = c$, then

$$g_{trt}(\theta) = -G\,\frac{I_E}{r_E^{\,4}}\,\frac{v_{Eq}}{c}\left(\frac{\Omega_\phi(\theta) - \Omega_E}{\Omega_E}\right)\cos^2(\lambda(\theta))PS(r(\theta)) \quad , \tag{6.10}$$

where the formula for $PS(r)$ is given by Eq. (2.43). Calculated values for $g_{trt}$ are

$$g_{trt}(0) = +11.199 \times 10^{-12}\ \text{m/s}^2 \quad ,$$
$$g_{trt}(\pm\pi) = +8.093 \times 10^{-12}\ \text{m/s}^2 \quad . \tag{6.11}$$

The formula for the transverse induction-like field is given by Eq. (3.8).

$$F_\lambda(\theta) = \frac{v_{Eq}}{v_k}\,\frac{r_E}{rM(\theta)}\int_0^\theta \frac{rM(\theta)}{r_E}\,\frac{\Omega_0(\theta)}{\Omega_E}\,\frac{1}{r_E}\,\frac{drM}{d\theta}\,\frac{dg_{trt}}{d\theta}\,d\theta \quad . \tag{6.12}$$

The derivative $dg_{trt}/d\theta$ can be found by using numerical differentiation. Let's start with the following trial value for the induction speed.

$$v_k = 7.322 v_{Eq} \quad . \tag{6.13}$$

Numerical integration of Eq. (6.12) gave the following values for $F_\lambda(\theta)$.

$$F_\lambda(0) = 0 \quad ,$$
$$F_\lambda(-\pi) = +4.2164 \times 10^{-14}\ \text{m/s}^2 \quad ,$$
$$F_\lambda(+\pi) = -4.2164 \times 10^{-14}\ \text{m/s}^2 \quad . \tag{6.14}$$

The formula for the speed change is given by Eq. (4.3). For this case, $v_{in} = vM_a$.

$$\delta v(\theta) = \frac{vM_a}{2}\int_0^\theta \frac{r_\lambda(\theta)F_\lambda(\theta)}{vM_a^{\,2}}\,\frac{d\lambda}{d\theta}\,d\theta \quad . \tag{6.15}$$

The derivative $d\lambda/d\theta$ can be found by numerical differentiation of Eq. (6.7). Values for $\delta v(\theta)$ can be found by numerical integration of Eq. (6.15).

The formula for the total speed change per revolution is given by Eqs. (4.4) and (4.5). The value for $\theta_{min}=-\pi$ and for $\theta_{max}=+\pi$.

$$\delta v_{in} = \delta v(\theta_{min}) = -1.3343 \times 10^{-9} \text{ m/s} \ ,$$
$$\delta v_{out} = \delta v(\theta_{max}) = -1.3343 \times 10^{-9} \text{ m/s} \ ,$$
$$\delta v_{trt} = \delta v_{in} + \delta v_{out} = -2.6687 \times 10^{-9} \text{ m/s per revolution} \ . \tag{6.16}$$

Let $N_{rev}$ be the number of lunar revolutions per year, and let $\delta vM$ be the total orbital speed change per year. Then ($P_{sol}$ is from Appendix B)

$$P_{sol} = 29.530589 \text{ days} \ ,$$
$$N_{rev} = \frac{365.25}{P_{sol}} = 12.3685 \ ,$$
$$\delta vM = N_{rev}\delta v_{trt} = -33.01 \times 10^{-9} \text{ m/s per year} \ . \tag{6.17}$$

Thus we find that, with $v_k=7.322v_{Eq}$, the calculated value for the Moon's orbital speed-change is $-33\times10^{-9}$ m/s per year, which explains exactly the lunar orbit anomaly.

Uncertainty in the required value for $v_k$ is mainly due to uncertainty in the inclination of the Moon's orbit (*cf* Eq. (6.6)). The corresponding uncertainty in $v_k/v_{Eq}$ is $\pm1$. The final value for $v_k/v_{Eq}$ for the Moon becomes

$$\frac{v_k}{v_{Eq}} = 7 \pm 1 \ . \tag{6.18}$$

## 7. Conclusions and recommendations

There is here within conclusive evidence that the neoclassical causal version of Newton's acausal theory agrees with the facts-of-observation to the extent that such facts are currently available. There were skeptics who initially refused to accept Newton's theory, but as new facts-of-observation were accumulated, Newton's theory was shown to explain them and it became widely accepted. The causal theory proposed herein, which is a natural rational extension of Newton's acausal theory to a causal version of the theory, can be expected to apply only for slow-speeds and weak-fields. It is recommended that various available methods be used to detect the effects of time retardation.

## Acknowledgements

I thank Patrick L. Ivers for reviewing the manuscript and suggesting improvements. I also thank Dr. Robert A. Nelson for bringing to my attention the study of records of eclipses by F. R. Stephenson and L. V. Morrison.[3]

## Appendix A: Numerical values

Most of the following values are readily available in the literature.

| | |
|---|---|
| $G = 6.6732 \times 10^{-11} \ \dfrac{m^3}{kg \cdot s^2}$ | Gravity constant |
| $c = 2.997925 \times 10^8 \ m/s$ | Vacuum speed of light |
| $\Omega_E = 7.292115 \times 10^{-5} \ rad/s$ | Earth's sidereal angular speed |
| $M_E = (5.9761 \pm 0.004) \times 10^{24} \ kg$ | Earth's total mass |
| $r_E = (r_e^2 r_p)^{\frac{1}{3}} = 6 \ 371 \ 034 \pm 21 \ m$ | Earth's equivalent spherical radius |
| $v_{Eq} = r_E \Omega_E = 464.58 \ m/s$ | Earth's equatorial surface speed |
| $V_E = (1.08322 \pm 0.00001) \times 10^{21} \ m^3$ | Earth's volume |
| $\bar{\rho}_E = (5.517 \pm 0.004) \times 10^3 \ kg/m^3$ | Earth's mean mass-density |
| $I_E = (8.0238 \pm 0.0085) \times 10^{37} \ kg \cdot m^2$ | Earth's spherical moment of inertia |

## Appendix B: Numerical values for the Moon

The following numerical values can be found in the article "Moon" in the Wikipedia Encyclopedia.[8]

| | |
|---|---|
| $r_M = 1737.10 \ km$ | Moon's mean radius |
| $V_M = 2.1958 \times 10^{19} \ m^3$ | Moon's volume |
| $M_M = 7.3477 \times 10^{22} \ kg$ | Moon's mass |
| $r_p = 363104 \times 10^3 \ m$ | Radial distance at perigee |
| $r_a = 405696 \times 10^3 \ m$ | Radial distance at apogee |
| $\varepsilon = 0.0549$ | Eccentricity |
| $P_{sid} = 27.321582 \ days$ | Sidereal orbital period |
| $P_{sol} = 29.530589 \ days$ | Solar orbital period |
| $\alpha_{Eq} = (18.29° + 28.58°)/2$ | Average inclination to Earth's |
| $\quad = 23.43° \pm 5.15°$ | equatorial plane |

––––––––––––––––––––––

1. Retired PhD Professor of Physics, Math, and Computer Science. Home office: 618 S. 24$^{th}$ St., Laramie, WY 82070, USA. Email: cahafele@bresnan.net.